\documentstyle[amstex,aps,epsfig,graphicx]{revtex}

\begin{document}
\title{Spontaneous pattern formation in an acoustical resonator}
\author{V.J. S\'{a}nchez-Morcillo}
\address{Departamento de F\'{i}sica Aplicada, Universidad Polit\'{e}cnica de\\
Val\`{e}ncia,\\
Carretera Nazaret-Oliva S/N, 46730 Grao de Gandia (Spain)}
\maketitle

\begin{abstract}
A dynamical system of equations describing parametric sound generation (PSG)
in a dispersive large aspect ratio resonator is derived. The model
generalizes previously proposed descriptions of PSG by including diffraction
effects, and is analogous to the model used in theoretical studies of
optical parametric oscillation.

A linear stability analysis of the solution below the threshold of
subharmonic generation reveals the existence of a pattern forming
instability, which is confirmed by numerical integration. The conditions of
emergence of periodic patterns in transverse space are discussed in the
acoustical context.
\end{abstract}

\section{Introduction.}

In recent years, pattern formation in systems driven far from equilibrium
has became an active field of research in many areas of nonlinear science.
Apart from peculiarities of particular systems, an outstanding property of
pattern formation is its universality, evidenced when the dynamical models
describing the different phenomena (either hydrodynamical, chemical, optical
or others) can be reduced, under several approximations, to the same order
parameter equation. These equations are few and well known, such as
Ginzburg-Landau or Swift Hohenberg \cite{Cross93}.

A key concept in pattern forming systems is the aspect ratio. When the
evolution of the variables is restricted to a bounded region of space, or
cavity, the aspect ratio is defined as the ratio of transverse to
longitudinal sizes of the cavity. In hydrodynamical Rayleigh-B\`{e}nard
convection, for example, the aspect ratio is determined by the ratio between
the height of the fluid and the area of the container. In problems of
nonlinear wave interaction, this parameter is related to the Fresnel number,
usually defined as 
\begin{equation}
F=\frac{a^{2}}{\lambda L},  \label{fresnel}
\end{equation}
where $a$ is the characteristic transverse size of the cavity (for example,
the area of a plane radiator), $\lambda $ is the wavelength and $L$ is the
length of the cavity in the direction of propagation, considered the
longitudinal axis of the cavity.

Spontaneous pattern formation is observed in large aspect ratio nonlinear
systems driven by an external input, where the possibility of excitation of
many transverse modes (a continuum for an infinite transverse dimension) is
considered. When the amplitude of the external input reaches a critical
threshold value, large enough to overcome the losses produced by dissipative
processes in the system, a symmetry breaking transition occurs, carrying the
system from an initially homogeneous to a inhomogeneous state, usually with
spatial periodicity. These solutions have been often called dissipative
structures \cite{Prigogine67}. A paradigmatic example in pattern formation
studies has been the Rayleigh-B\`{e}nard convection in a fluid layer heated
from below, where roll or hexagonal patterns are excited above a given
temperature threshold \cite{Cross93}.

This scenario differs with that observed in small aspect ratio systems such
as, for example, a waveguide resonator of finite cross section. In this
case, boundary-induced spatial patterns are selected not by the nonlinear
properties of the cavity, but by the transverse boundary conditions, and
correspond the excitation of one or few transverse modes of the cavity \cite
{Ostrovsky76}. In this sense, boundary-induced patterns are of linear nature.

Guided by the analogies with other physical systems, we can expect that an
ideal system for such effects to be observed in acoustics consists in a
resonator of plane walls (acoustical interferometer) with infinite
transverse size. In practice, the large aspect ratio condition could be
fulfilled if the pumped area is finite, but large in comparison with the
spatial scale imposed by the cavity length and the field wavelength, as
follows from (\ref{fresnel}).

Parametrically driven systems offer many examples of spontaneous pattern
formation. For example, parametric excitation of surface waves by a vertical
excitation (Faraday instability)\cite{Miles84}, vibrated granular layers 
\cite{Melo94}, spin waves in ferrites and ferromagnets and Langmuir waves in
plasmas parametrically driven by a microwave fields \cite{Lvov94}, or the
optical parametric oscillator \cite{Oppo94,deValcarcel96} have been studied.

The behaviour of nonlinear waves in large aspect ratio resonators has been
extensively studied in nonlinear optics (for a recent review, see \cite
{Arecchi99}), where a rich variety of patterns has been observed. On the
other side, it is well known that optical and acoustical waves share many
common phenomena, under restricted validity conditions \cite{Bunkin86}.

In nonlinear acoustics, a phenomenon belonging to the class of the previous
examples is the parametric sound amplification. It consists in the resonant
interaction of a triad of waves with frequencies $\omega _{0},\omega _{1}$
and $\omega _{2}$, for which the following energy and momentum conservation
conditions are fulfilled: 
\begin{eqnarray}
\omega _{0} &=&\omega _{1}+\omega _{2},  \nonumber \\
\vec{k}_{0} &=&\vec{k}_{1}+\vec{k}_{2}+\Delta \vec{k},  \label{conservation}
\end{eqnarray}
where $\Delta \vec{k}$ is a small phase mismatch. The process is initiated
by an input pumping wave of frequency $\omega _{0}$ which, due to the
propagation in the nonlinear medium, generates a pair of waves with
frequencies $\omega _{1}$ and $\omega _{2}$. When the wave interaction
occurs in a resonator, a threshold value for the input amplitude is
required, and the process is called parametric sound generation. In
acoustics, this process has been described before by several authors under
different conditions, either theoretical and experimentally. In \cite
{Korpel65,Adler70,Yen75}, the one dimensional case (colinearly propagating
waves) is considered. In \cite{Ostrovsky76}, the problem of interaction
between concrete resonator modes, with a given transverse structure, is
studied. In both cases, small aspect ratio resonators containing liquid and
gas respectively are considered. More recently, parametric interaction in a
large aspect ratio resonator filled with superfluid He$^{4}$ has been
investigated \cite{Rinberg96,Rinberg01}.

The phenomenon of parametric sound generation is analogous to optical
parametric oscillation in nonlinear optics. However, an important difference
between acoustics and optics is the absence of dispersion in the former.
Dispersion, which makes the phase velocity of the waves to be dependent on
its frequency, allows that only few waves, those satisfying given
synchronism conditions, participate in the process.

In a nondispersive medium, all the harmonics of each initial monochromatic
wave propagate synchronously. As a consequence, the spectrum broadens during
propagation and the energy is continuously pumped into the higher harmonics,
which eventually leads to shock formation.

Optical media are in general dispersive, but acoustical media not. In finite
geometries, such as waveguides\cite{Hamilton87} or resonators\cite
{Ostrovsky78}, the dispersion is introduced by the boundaries. Different
cavity modes propagate at different angles, and then with different
''effective'' phase velocities. However, in unbounded systems
boundary-induced dispersion is not present.

Other dispersion mechanisms have been proposed in nonlinear acoustics, such
as bubbly media or layered (periodic) media \cite{Hamiltonbook}. In all this
systems, dispersion appears due to the introduction of additional spatial or
temporal scales in the system, which makes sound velocity propagation to be
wavelength dependent. Other proposed mechanisms are, for example, media with
selective absorption, in which selected spectral components experience
strong losses and may removed from the wave field\cite{Zarembo74}.

Pattern formation in acoustics has been reported previously in the context
of acoustic cavitation, \cite{Akhatov94,Akhatov96}, where the coupling
between the sound field amplitude and the bubble distribution is considered.
We do not consider here the dynamics of the medium, which is assumed to be
at rest, but the coupling between different frequency components.

The aim of the paper is twofold. On one side, a rigurous derivation of the
dynamical model describing parametric interaction of acoustic waves in a
large aspect ratio cavity is presented. The derived model is analogous to
the system of equations describing parametric oscillation in an optical
resonator, and consequently their solutions are known. On the other side,
among other peculiarities, the model predicts a pattern forming instability,
which is confirmed by a numerical integration. We review these properties in
the acoustical context, giving some estimations of the acoustical parameters
which can make the model closer to real operating conditions.

\section{Model equations.}

The starting point of the theoretical analysis is the nonlinear wave
equation, which written in terms of the pressure $p$ reads 
\begin{equation}
\nabla ^{2}p-\frac{1}{c^{2}}\frac{\partial ^{2}p}{\partial t^{2}}+\frac{%
\delta }{c^{4}}\frac{\partial ^{3}p}{\partial t^{3}}=-\frac{\varepsilon }{%
\rho _{0}c^{4}}\frac{\partial ^{2}p^{2}}{\partial t^{2}}-\left( \nabla ^{2}+%
\frac{1}{c^{2}}\frac{\partial ^{2}}{\partial t^{2}}\right) {\cal L},
\label{nwe}
\end{equation}
where ${\cal L}=\frac{1}{2}\rho _{0}v^{2}-\frac{p^{2}}{2\rho _{0}c^{2}}$ is
the lagrangian density, with $\rho _{0}$ the ambient pressure and $v$ the
particle velocity, $\delta $ is the diffusivity of sound, defined as 
\begin{equation}
\delta =\frac{1}{\rho _{0}}\left( \frac{4}{3}\eta +\zeta +\kappa \left( 
\frac{1}{C_{v}}-\frac{1}{C_{p}}\right) \right) ,  \label{diffusivity}
\end{equation}
and $\varepsilon =1+B/2A$ is the nonlinearity coefficient.

In the case of colinearly propagating plane waves, the lagrangian density
term vanishes, since the linear impedance relation $p=\rho _{0}cv$ holds.
However, even in the case of slightly diverging waves (plane waves
propagating at a small angle), the last term in (\ref{nwe}) is much smaller
in magnitude than the other terms in the equation, and in this approximation
its effects can be ignored \cite{Hamilton87}.

Furthermore, in this case the nonlinearity parameter can be considered to be
independent of the interaction angle. It has been shown in \cite{Hamilton87}
that, for the process described by Eqs.(\ref{conservation}), when the waves
with frequencies $\omega _{1}$ and $\omega _{2}$ propagate at an angle $%
\theta $, the nonlinearity parameter for the $\omega _{0}\,$wave is
expressed as 
\begin{equation}
\varepsilon _{0}\left( \theta \right) =\cos \theta +4\frac{\omega _{1}\omega
_{2}}{\omega _{0}^{2}}\sin ^{4}\left( \frac{\theta }{2}\right) +\frac{B}{2A}
\label{nonlinearity}
\end{equation}
which reduces to $\varepsilon =1+B/2A$ when $\theta $ is small.

Under this assumptions, the field distribution can be properly described by
the wave equation 
\begin{equation}
\nabla ^{2}p-\frac{1}{c^{2}}\frac{\partial ^{2}p}{\partial t^{2}}-\frac{%
\delta }{c^{4}}\frac{\partial ^{3}p}{\partial t^{3}}=-\frac{\varepsilon }{%
\rho _{0}c^{4}}\frac{\partial ^{2}p^{2}}{\partial t^{2}},  \label{westervelt}
\end{equation}
which is the well known Westervelt equation.

Equation (\ref{westervelt}) describes the propagation of waves in a
nondispersive homogeneous medium, but also in a medium that possess some
dispersion mechanism, such as a bubbly liquid when the field frequencies are
much lower than the resonance frequency of the bubbles \cite{Druzhinin96}.
It must be noted that, in a dispersive medium, the nonlinearity parameter $%
\varepsilon $ depends on the dispersion mechanism.

We consider in the following that the wave interaction is only effective
among three resonant frequencies, for which the relations (\ref{conservation}%
) hold. Taking into account that, due to reflections in the walls, there
exist waves propagating simultaneously in opposite directions, the field
inside the resonator can then be expanded as 
\begin{equation}
p({\bf r},t)=\sum_{j=1}^{3}P_{j}({\bf r},t),  \label{sum}
\end{equation}
where $P_{j}({\bf r},t)$ are the wave components related with the frequency $%
\omega _{j}$, given by 
\begin{equation}
P_{j}({\bf r},t)=p_{j}(x,y,t)\cos \left( k_{j}^{c}z\right) e^{-i\omega
_{j}t}+c.c.  \label{cosmodes}
\end{equation}
where $k_{j}^{c}=m\pi /L$ is a cavity eigenmode.

A more general solution can be proposed, consisting in a superposition of
quasi-planar waves, the field being decomposed in forward ($F$) and backward
($B$) waves, respectively. The quasi-planar assumption implies that the
amplitudes may depend on the longitudinal coordinate $z$, representing a
slow evolution compared with the scale given by $k_{j}.$ The field at
frequency $\omega _{j}$ is then expressed as 
\begin{equation}
P_{j}({\bf r},t)=\left[ F_{j}({\bf r},t)e^{ik_{j}z}+B_{j}({\bf r}%
,t)e^{-ik_{j}z}\right] e^{-i\omega _{j}t}+c.c.  \label{fbmodes}
\end{equation}
where ${\bf r=}\left( x,y,z\right) .$

The case $\omega _{0}=2\omega _{1}$ corresponds to degenerate interaction ($%
\omega _{1}=\omega _{2}=\omega _{0}/2$), and describes the process of second
harmonic generation or subharmonic parametric generation, depending on
whether the pumping wave oscillates at $\omega _{0}$ or $\omega _{1.}$ In
the following, the degenerate interaction case is considered, where 
\begin{equation}
p({\bf r},t)=P_{0}\left( {\bf r},t\right) +P_{1}\left( {\bf r},t\right)
\label{twofields}
\end{equation}
with the amplitudes given in (\ref{fbmodes}).

Substitution of (\ref{twofields}) in (\ref{westervelt}), and projecting the
resulting equation on each of the mode frequencies, two coupled wave
equations are found, 
\begin{equation}
\nabla ^{2}P_{j}-\frac{1}{c_{j}^{2}}\frac{\partial ^{2}P_{j}}{\partial t^{2}}%
+\frac{\delta _{j}}{c_{j}^{4}}\frac{\partial ^{3}P_{j}}{\partial t^{3}}=-%
\frac{\varepsilon }{\rho _{0}c_{j}^{4}}\frac{\partial ^{2}\left(
p^{2}\right) _{j}}{\partial t^{2}}  \label{westmodes}
\end{equation}
where $j=0,1$ and $c_{j}=\omega _{j}/k_{j}=c\left( \omega _{j}\right) $ is
the phase velocity of the waves.

The evolution equations for the amplitudes $p_{j}$ at the different
frequencies can derived be from (\ref{westmodes}) under several
approximations. The complete derivation is given in the Appendix. If we
assume that ($i$) the amplitudes are slowly varying in $z$ and $t$, ($ii$)
the reflectivity ${\cal R}_{j}$ at the boundaries is high, ($iii$) the field
frequencies $\omega _{j}$ are close to one resonator frequency, $\omega
_{j}^{c}$ and that ($iv$) only one longitudinal mode is excited by each
frequency component, the evolution is ruled by the following dynamical
equations:

\begin{eqnarray}
\frac{\partial p_{0}}{\partial t} &=&E-\gamma _{0}\left( 1+i\Delta
_{0}\right) p_{0}+ia_{0}\nabla _{\perp }^{2}p_{0}-i\frac{b_{0}}{4}p_{1}^{2},
\nonumber \\
\frac{\partial p_{1}}{\partial t} &=&-\gamma _{1}\left( 1+i\Delta
_{1}\right) p_{1}+ia_{1}\nabla _{\perp }^{2}p_{1}-i\frac{b_{1}}{2}%
p_{1}^{\ast }p_{0}  \label{psg1}
\end{eqnarray}
where $p_{j}\left( x,y,t\right) $ are the amplitudes corresponding to (\ref
{cosmodes}), the nonlinearity parameter is defined as $b_{j}=\varepsilon
k_{j}/2\rho _{0}c_{j}$, and the asterisk denoted complex conjugation$.\,$The
other parameters in (\ref{psg1}) are the pump $E$, detuning $\Delta _{j}$,
losses $\gamma _{j}$\ and diffraction $a_{j}$ and are defined as 
\begin{eqnarray}
E &=&\frac{1}{2L}\frac{\sqrt{{\cal T}_{0}}}{\sqrt{{\cal R}_{0}}}Ye^{i\frac{%
\theta _{0}}{2}},  \label{pump} \\
\Delta _{j} &=&\frac{\theta _{j}}{\left| \ln {\cal R}_{j}\right| +\frac{2L}{%
c_{j}}\kappa _{j}},  \label{detuning} \\
\gamma _{j} &=&\frac{\left| \ln {\cal R}_{j}\right| c_{j}}{2L}+\kappa _{j},
\label{gamma} \\
a_{j} &=&\frac{c_{j}}{2k_{j}}  \label{diffrac}
\end{eqnarray}
where $Y$ is the amplitude of the incident wave at frequency $\omega _{0}$, $%
{\cal R}_{j}$ are the reflectivities at the boundaries, $\theta
_{j}=2L\left( \omega _{j}^{c}-\omega _{j}\right) /c_{j}$ is the frequency
mismatch, and $\kappa _{j}=\delta k_{j}^{2}/2$ is a loss factor related with
the diffusivity of sound. Note that the total loss parameter $\gamma _{j}$
takes into account both loss mechanisms, due to absorption and reflectivity
at the boundaries (energy leakage of the resonator). Also, from (\ref
{detuning}) and (\ref{gamma}) the detuning can be written as $\Delta
_{j}=\left( \omega _{j}^{c}-\omega _{j}\right) /\gamma _{j}.$

Finally, the field amplitudes are normalized to leave the model in its final
form, 
\begin{eqnarray}
\frac{1}{\gamma _{0}}\frac{\partial A_{0}}{\partial t} &=&{\cal E}%
-(1+i\Delta _{0})A_{0}+ia_{0}\nabla ^{2}A_{0}-A_{1}^{2},  \nonumber \\
\frac{1}{\gamma _{1}}\frac{\partial A_{1}}{\partial t} &=&-(1+i\Delta
_{1})A_{1}+ia_{1}\nabla ^{2}A_{1}+A_{0}A_{1}^{\ast }.  \label{psg2}
\end{eqnarray}
where $A_{0}=i(b_{1}/2)p_{0},\,A_{1}=i\sqrt{b_{0}b_{1}/8}p_{1}$ and ${\cal E}%
=i(b_{1}/2)E$.

The system of equations (\ref{psg2}), together with their complex
conjugated, consists in the model for parametric interaction of acoustic
waves in large aspect ratio resonators. Eqs.(\ref{psg2}) are suitable for
the description of the spatio-temporal evolution of the pressure envelope
waves oscillating at the fundamental $\left( \omega _{0}\right) $ and
subharmonic $\left( \omega _{1}\right) $ frequencies.

These equations have been studied in the context of nonlinear optics, as a
model for optical parametric oscillation \cite{Oppo94}. In the following we
review some of the basic properties of the solutions of Eqs.(\ref{psg2}),
and their application to the case of the acoustic resonator.

\section{Homogeneous solutions}

The stationary and spatially homogeneous solutions of (\ref{psg2}) are
obtained when the temporal derivatives and the transverse diffraction term
vanish. In this case, the model reduces to the one derived for parametric
sound generation in a resonator with rectangular cross section $a\times b$,
described in \cite{Ostrovsky76}, where its homogeneous solutions were
obtained. We next review these solutions in the present notation.

The simplest homogeneous solution corresponds to the trivial solution, 
\begin{equation}
\bar{A}_{0}=\frac{{\cal E}}{1+i\Delta _{0}},\,\bar{A}_{1}=0,  \label{trivial}
\end{equation}
characterized by a null value of the subharmonic field inside the resonator.
This solution exists for low values of the pump amplitude (below the
instability threshold to be discussed in the next section).

For larger pump values, also the subharmonic field has a nonzero amplitude,
given by 
\begin{equation}
\left| \bar{A}_{1}\right| ^{2}=-1+\Delta _{0}\Delta _{1}\pm \sqrt{{\cal E}%
^{2}-\left( \Delta _{1}+\Delta _{1}\right) ^{2}},  \label{notriv1}
\end{equation}
while the stationary fundamental intensity $\left| A_{0}\right| ^{2}$ takes
the value 
\begin{equation}
\left| \bar{A}_{0}\right| ^{2}=1+\Delta _{1}^{2},  \label{notriv0}
\end{equation}
which is independent of the value of the injected pump.

The emergence of this finite amplitude solution corresponds to the process
of parametric generation. In the frame of the plane wave model it has been
shown theoretically, and confirmed experimentally \cite{Ostrovsky76}, that
the trivial solution (\ref{trivial}) bifurcates in the nontrivial one, and
the subharmonic field emerges with an amplitude given by (\ref{notriv1}).
This bifurcation is supercritical when $\Delta _{0}\Delta _{1}<1$, and
subcritical when $\Delta _{0}\Delta _{1}>1$. In the latter case, both
homogeneous solutions can coexist for given sets of the parameters.

In the next section this scenario is generalized by including diffraction
effects in the model.

\section{Linear stability analysis}

In order to study the stability of the trivial solution (\ref{trivial})
against space-dependent perturbations, consider a deviation of this state,
given by 
\begin{equation}
A_{j}\left( x,y,t\right) =\bar{A}_{j}+\delta A_{j}\left( x,y,t\right) .
\label{deviation}
\end{equation}

Assuming the deviation to be small, after substitution of (\ref{deviation})
in (\ref{psg2}) the resulting system can be linearized in the perturbations $%
\delta A_{j}$. This leads to the linear system of equations 
\begin{eqnarray}
\frac{1}{\gamma _{0}}\frac{\partial \delta A_{0}}{\partial t} &=&-(1+i\Delta
_{0})\delta A_{0}+ia_{0}\nabla ^{2}\delta A_{0},  \nonumber \\
\frac{1}{\gamma _{1}}\frac{\partial \delta A_{1}}{\partial t} &=&-(1+i\Delta
_{1})\delta A_{1}+ia_{1}\nabla ^{2}\delta A_{1}-\bar{A}_{0}\delta
A_{1}^{\ast }.  \label{lin2}
\end{eqnarray}

The generic solutions of (\ref{lin2}) are of the form 
\begin{equation}
\left( \delta A_{j},\delta A_{j}^{\ast }\right) \propto e^{\lambda \left(
k_{\perp }\right) t}e^{i{\bf k}_{\perp }\cdot {\bf r}_{\perp }},
\label{perturb}
\end{equation}
where $\lambda ({\bf k}_{{\bf \perp }})$ represents the growth rate of the
perturbations, and ${\bf k}_{{\bf \perp }}$ is the transverse component of
the wavevector, which in a two-dimensional geometry obeys the relation $%
\left| {\bf k}_{{\bf \perp }}\right| ^{2}=k_{x}^{2}+k_{y}^{2}.$

Substitution of (\ref{perturb}) in (\ref{lin2}) and its complex conjugates,
written in matrix form, allows to evaluate the growth rates $\lambda $ as
the eingenvalues of the stability matrix $L$. This is given by 
\begin{equation}
L=\left( 
\begin{array}{cc}
L_{0}\left( k_{\perp }\right)  & 
\begin{array}{cc}
0 & 0 \\ 
0 & 0
\end{array}
\\ 
\begin{array}{cc}
0 & 0 \\ 
0 & 0
\end{array}
& L_{1}\left( k_{\perp }\right) 
\end{array}
\right) ,  \label{L}
\end{equation}
where the block matrices are defined as 
\begin{equation}
L_{0}\left( k_{\perp }\right) =\gamma _{0}\left( 
\begin{array}{cc}
-1+i\Delta _{0}+ia_{0}k_{\perp }^{2} & 0 \\ 
0 & -1-i\Delta _{0}-ia_{0}k_{\perp }^{2}
\end{array}
\right) ,  \label{L0}
\end{equation}
and 
\begin{equation}
L_{1}\left( k_{\perp }\right) =\gamma _{1}\left( 
\begin{array}{cc}
-1+i\Delta _{1}+ia_{1}k_{\perp }^{2} & \frac{{\cal E}}{1-i\Delta _{0}} \\ 
\frac{{\cal E}}{1+i\Delta _{0}} & -1-i\Delta _{1}-ia_{1}k_{\perp }^{2}
\end{array}
\right) .  \label{L1}
\end{equation}

The eigenvalues of $L_{0}$ have always a negative real part, as follows from
(\ref{L0}), and do not predict any instability. The eigenvalues of $L_{1}$
are associated with the instability which gives rise to the subharmonic
field, and are given by 
\begin{equation}
\lambda _{\pm }\left( k_{\perp }\right) =-1\pm \sqrt{\frac{{\cal E}^{2}}{%
1+\Delta _{0}^{2}}-\left( \Delta _{1}+a_{1}k_{\perp }^{2}\right) ^{2}}.
\label{eigenvalue}
\end{equation}

Note that only the eigenvalue with the positive sign $\lambda _{+}$ can take
positive values, and reach the instability condition $%
%TCIMACRO{\func{Re}}%
%BeginExpansion
\mathop{\rm Re}%
%EndExpansion
\left( \lambda \right) >0.$

The eigenvalue is wavenumber dependent, which means that not all the
perturbations in the form of transverse plane-wave modes (\ref{perturb})
grow at the same rate. The maximum growth rate follows from the condition $%
\partial \lambda _{+}/\partial k=0$. Two different cases, depending on the
sign of the subharmonic detuning, must be considered:

If $\Delta _{1}>0$, which corresponds to a subharmonic frequency larger than
that of the closest cavity mode, the eigenvalue shows a maximum at 
\begin{equation}
k_{\perp }=0.  \label{khomog}
\end{equation}

In this case, the emmited subharmonic wave travels parallel to the cavity
axis, without spatial variations on the transverse plane. The solution
remains homogeneous, and its amplitude is given by (\ref{notriv1}).

On the contrary, if $\Delta _{1}<0$, corresponding to a field frequency
smaller than that of the cavity mode, the maximum of the eigenvalue occurs
at 
\begin{equation}
k_{\perp }=\sqrt{-\frac{\Delta _{1}}{a_{1}}}.  \label{kpattern}
\end{equation}

The corresponding solution is of the form (\ref{perturb}), which represents
a plane wave tilted with respect to the cavity axis. This solution presents
spatial variations in the transverse plane, and consequently pattern
formation is expected to occur. The two kind of instabilities are
represented in Fig.1, where the eigenvalue as a function of the wavenumber
is plotted, in the cases of positive ($a$) and negative ($b$) detuning, for
given values of the pump above threshold.

The value of the transverse wavenumber given by (\ref{kpattern}) can be
interpreted in simple geometrical terms: $k_{\perp }$ corresponds to the
tilt of the wave necessary to fit the longitudinal resonance condition. This
is a linear effect related with diffraction, and it is represented in the
figure 2.

This can be also analytically shown by inspecting the solutions of the
dynamical equations for a transverse wave in the form (\ref{perturb}): the
frequency of the cavity mode is given by $\omega =c\left| {\bf k}\right| =c%
\sqrt{k_{\perp }^{2}+k_{z}^{2}}$. Since, in the small detuning case, the
relation $k_{\perp }<<k_{z}$ holds, it can be written approximately by 
\begin{equation}
\omega =ck_{z}\sqrt{1+\left( \frac{k_{\perp }}{k_{z}}\right) ^{2}}\approx
ck_{z}+\frac{c}{2k_{z}}k_{\perp }^{2}=ck_{z}+\Delta \omega
\end{equation}
where $\Delta \omega $ is the transverse contribution to the mode frequency.
This contribution arises from the diffraction term $ia\nabla ^{2}A=i\frac{c}{%
2k_{z}}\nabla ^{2}A$ in the dynamical equations.

Since $k_{\perp }$ is the modulus of the wavevector, the linear stability
analysis in two dimensions predicts that a continuum of modes within a
circular annulus (centered on a critical circle at $\left| {\bf k}_{\perp
}\right| =k_{\perp }$ in $\left( k_{x},k_{y}\right) $ space) grows
simultaneously as the pump increases above a critical value. This double
infinite degeneracy of spatial modes (degenerate along a radial line from
the origin and orientational degeneracy) allows, in principle, arbitrary
structures in two dimensions.

The threshold (the pump value at which the instability emerges) depends also
on the sign of the detuning, and is obtained from the condition $%
%TCIMACRO{\func{Re}}%
%BeginExpansion
\mathop{\rm Re}%
%EndExpansion
\lambda =0$. From (\ref{eigenvalue}) it follows that 
\begin{equation}
{\cal E}_{th}=\sqrt{1+\Delta _{0}^{2}}\sqrt{1+\left( \Delta
_{1}+a_{1}k^{2}\right) ^{2}}.  \label{threshold1}
\end{equation}

From the previous analysis, again two cases must be distinguished. For
positive detunings, the mode homogeneous $k_{\perp }=0$ experience the
maximum growth, and the threshold occurs at a pump value 
\begin{equation}
{\cal E}_{th}=\sqrt{1+\Delta _{0}^{2}}\sqrt{1+\Delta _{1}^{2}},
\label{threshold2}
\end{equation}
which is the same found in \cite{Ostrovsky78}. The solution above the
threshold is given by (\ref{notriv1}).

For negative detunings, the instability leads to a non homogeneous
distribution, with a characteristic scale given by the condition $%
a_{1}k_{\perp }^{2}=-\Delta _{1}$. The corresponding threshold for such
modes is 
\begin{equation}
{\cal E}_{th}=\sqrt{1+\Delta _{0}^{2}}.  \label{threshold3}
\end{equation}

Besides the existence of a pattern forming instability, a relevant
conclusion of the previous analysis is the prediction of a decrease in the
threshold pump value of subharmonic generation when diffraction effects are
included [compare Eqs.(\ref{threshold2}) and (\ref{threshold3})]. This fact
is specially important in acoustical systems where the nonlinearity is weak,
since in this case the instability threshold, appearing for high values of
the driving, can be lowered by a factor of $\sqrt{1+\Delta _{1}^{2}}$.

The predictions of the stability analysis correspond to the linear stage of
the evolution, where the subharmonic field amplitude is small enough to be
considered a perturbation of the trivial state. The analytical study of the
further evolution would require a nonlinear stability analysis, not given
here. Instead, in the next section we perform the numerical integration of
Eqs.(\ref{psg2}), where predictions of the acoustic subharmonic\ field in
the linear and nonlinear regime are given.

\section{Numerical simulations}

In order to check the analytical predictions of the linear stability
analysis, we integrated numerically the system (\ref{psg2}) by using the
split-step technique on a spatial grid of dimensions 64$\times $64. The
local terms, either linear (pump, losses and detuning) and nonlinear, are
calculated in the space domain, while nonlocal terms (diffractions) are
evaluated in the spatial wavevector (spectral) domain. A Fast Fourier
Transform (FFT) is used to shift from spatial to spectral domains in every
time step. Periodic boundary conditions are used.

As a initial contition, a noisy spatial distribution is considered, and the
parameters are such that a pattern forming instability is predicted (Fig.3a).

For small evolution times, the amplitude of the subharmonic remains small,
corresponding to the the linear stage of the evolution. As follows from the
linear stability analysis, an instability ring in the far field (transverse
wavenumber space) is observed (Fig.3b).

For larger evolution times, the nonlinearity comes into play. In the
nonlinear stage of evolution, a competition between transverse modes begins,
and pattern selection is observed. In Fig.4. it is shown a transient stage,
where a labyrinthic pattern is formed. The final state, not shown in the
figure, corresponds to the selection of a discrete set of transverse
wavevectors, asymptotically resulting in a periodic pattern.

\section{Acoustical estimates}

The theory presented in this paper has been suggested by analogies with
nonlinear optical resonators, where good agreement between theory and
experiment has been shown \cite{Arecchi99}. Next we estimate the required
physical conditions to make the theory applicable to an acoustical nonlinear
resonator.

The mean field assumption implies that the field envelopes changes little
during a roundtrip of the wave in the resonator. Acoustic waves, specially
in the ultrasound regime, experience strong losses during propagation in the
medium. It is then required the resonator to be short enough in order to
avoid strong absorption.

The longitudinal size of the resonator imposes a condition on the transverse
scale, in order to maintain the Fresnel number large. For example, for a
pump wave of 1 MHz in water the corresponding wavelength is $\lambda \approx
1500/10^{6}\approx 1.5$\ mm. If the resonator length is 2 cm and the walls
(transverse size of the transducer) are squared with 10 cm each side, the
Fresnel number gives $F=167,$ which can be considered in the large aspect
ratio limit.

With this geometry, the experimental setup proposed in the first observation
of parametric sound generation in a liquid filled resonator \cite{Korpel65}
seems a good candidate for the observation of the predicted phenomena. In
particular, it was shown in \cite{Korpel65} that, close to the threshold of
parametric generation, only the fundamental and subharmonic frequencies
where present inside the resonator, in agreement with our theory. The
resonator was formed by a PZT transducer, driven at a pumping frequency $%
\omega _{0}$, and a reflector placed at a variable distance $L$. The
threshold was achieved when the transducer was driven at a voltage of nearly 
$300$ $V$. The discrete spectrum, containing the fundamental $\omega $ and
the subharmonic $\omega /2$ frequencies, existed only close to the threshold
(the relative pumping level being 1dB). In some cases, pairs of signal-idler
waves (corresponding to the nondegenerate process) were excited. For larger
driving voltages, and even larger number of cavity modes, at different
combination frequencies, appeared due to the recurrent process of parametric
amplification. The evolution of the fields at such high pump values can not
be predicted by the model proposed in this paper, which was derived under
the assumption that the energy exchange occurs only between two modes.
Although the spectrum is still discrete in this case, the large number of
modes makes the spectral approach inconvenient for the theoretical study.
Instead, the field approach, typically used in problems of nonlinear
acoustic in nondispersive media, seems more convenient. A discussion
concerning the applicability of the two approaches to the description of
nonlinear wave processes in acoustics can be found in \cite{Rudenko95}.

\section{Conclusions}

In this paper, the problem of spontaneous emergence of patterns in an
acoustical interferometer is studied from the theoretical point of view. A
model for parametric sound generation in a large aspect-ratio cavity is
derived, taking into account the effects of diffraction. It is shown that
the subharmonic field can be excited, when a threshold pump value is
reached, characterized by a non-uniform distribution in the transverse
plane, or pattern. This occurs when the field frequency is tuned below the
frequency of the closest resonator mode (negative detuning). On the
contrary, an on-axis field, with homogeneous distribution, is emmited.
Traditionally, patterns arise from the imposition of external constraints
(waveguiding). The field then oscillates in modes of the resonator. The
spontaneous emergence of patterns considered in this article presupposes no
external transverse mode selection mechanism, but instead allows the system
to choose a pattern through the nonlinear interaction of a tipically
infinite set of degenerate modes. The pattern formation process described
here is related with the competing effects of nonlinearity and diffraction,
and presents many analogies with similar systems studied in nonlinear
optics, such as the two-level laser or the optical parametric oscillator, to
which the model derived in this paper is isomorphous.

\section{Acknowledgments}

The author thanks Dr. Y.N. Makov, V.E. Gusev, V. Espinosa, J. Ramis and J.
Alba for interesting discussions on the subject. The work was financially
supported by the CICYT of the Spanish Government, under the\ projects
PB98-0635-C03-02 and BFM2002-04369-C04-04.

\renewcommand{\theequation}{A-\arabic{equation}} \setcounter{equation}{0}

\section*{APPENDIX}

In this appendix we derive the model for parametric sound generation given
in (\ref{psg2}). The approach is based on the method proposed in \cite
{Lugiato88a} to describe the dynamics of laser fields in Fabry-Perot
resonators.

The first assumption is that the longitudinal variations of the fields are
mainly accounted for by the plane wave, in which case the envelope
amplitudes $F_{j}$ and $B_{j}$ can be considered as slowly varying functions
in $z$ and $t$. This leads to the inequalities 
\begin{equation}
\left| \frac{\partial ^{2}F_{j}}{\partial z^{2}}\right| <<\left| k_{j}\frac{%
\partial F_{j}}{\partial z}\right| ,\,\,\left| \frac{\partial ^{2}F_{j}}{%
\partial t^{2}}\right| <<\left| \omega _{j}\frac{\partial F_{j}}{\partial t}%
\right| .  \label{svea}
\end{equation}
which allow to neglect the second order derivatives in $z$ and $t$.

In the same way, applying the second relation of (\ref{svea}) to the
dissipation term, it reduces to 
\begin{equation}
\frac{\delta _{j}}{c_{j}^{4}}\frac{\partial ^{3}p_{j}}{\partial t^{3}}%
\approx i\frac{\delta _{j}}{c_{j}^{4}}\omega _{j}^{3}p_{j}
\label{dissipation}
\end{equation}

The condition (\ref{svea}) leads to the well known parabolic or eikonal
approximation, in which the d'Alembertian operator acting on the pressure on
the left-hand side of (\ref{westmodes}) can be approximated by 
\begin{equation}
\nabla ^{2}-\frac{1}{c_{j}^{2}}\frac{\partial ^{2}}{\partial t^{2}}\approx
2ik_{j}\left( \frac{\partial }{\partial z}+\frac{1}{c_{j}}\frac{\partial }{%
\partial t}\right) +\left( \frac{\partial ^{2}}{\partial x^{2}}+\frac{%
\partial ^{2}}{\partial y^{2}}\right) ,  \label{Eikonal}
\end{equation}

With (\ref{dissipation}) and (\ref{Eikonal}) the wave evolution can be
written as 
\begin{equation}
e^{ik_{j}z}\left( c_{j}\frac{\partial F_{j}}{\partial z}+\frac{\partial F_{j}%
}{\partial t}+\frac{c_{j}}{2ik_{j}}\nabla _{\bot }^{2}F_{j}-\kappa
_{j}F_{j}\right) +e^{-ik_{j}z}\left( -c_{j}\frac{\partial B_{j}}{\partial z}+%
\frac{\partial B_{j}}{\partial t}+\frac{c_{j}}{2ik_{j}}\nabla _{\bot
}^{2}B_{j}-\kappa _{j}B_{j}\right) =-ib_{j}\left[ p^{2}\right] _{j}^{-}
\end{equation}
where $\kappa _{j}=\frac{\delta _{j}k_{j}^{2}}{2}$ is a parameter that
accounts for dissipation, $b_{j}=\varepsilon k_{j}/2\rho _{0}c_{j}$ is a
nonlinearity parameter, and where $\left[ p^{2}\right] _{j}^{-}$ contains
the terms in $p^{2}$ that oscillate with frequency $-\omega _{j}$, and takes
into account slow and fast spatial variations.

The nonlinear terms at the frequencies $\omega _{0}$ and $\omega _{1}$ are,
respectively, 
\begin{eqnarray}
\left[ p^{2}\right] _{0}^{-} &=&\frac{1}{4}B_{1}^{2}e^{-ik_{0}z}+\frac{1}{2}%
B_{1}F_{1}+\frac{1}{4}F_{1}^{2}e^{ik_{0}z}  \nonumber \\
\left[ p^{2}\right] _{1}^{-} &=&\frac{1}{2}B_{1}^{\ast }B_{0}e^{-ik_{1}z}+%
\frac{1}{2}F_{1}^{\ast }B_{0}e^{-3ik_{1}z}+\frac{1}{2}B_{1}^{\ast
}F_{0}e^{3ik_{1}z}+\frac{1}{2}F_{1}^{\ast }F_{0}e^{ik_{1}z}
\label{nonlinear}
\end{eqnarray}

In order to eliminate the explicit dependence of the exponential factors in (%
\ref{waveamp}), a projection on two longitudinal modes is performed,
multiplying by $\frac{1}{2\pi }\exp \left( \pm ik_{j}z\right) $, and
integrating over a full wavelength. This leads to a system of equations
where the fields $F_{j}$ and $B_{j}$ are decoupled in the linear part ($%
l.h.s $), $i.e.$, we get 
\begin{equation}
c_{j}\frac{\partial F_{j}}{\partial z}+\frac{\partial F_{j}}{\partial t}+%
\frac{c_{j}}{2ik_{j}}\nabla _{\bot }^{2}F_{j}-\kappa _{j}F_{j}=-i\frac{b_{j}%
}{k_{j}}\frac{1}{2\pi }\int_{-\pi }^{\pi }\left[ p^{2}\right]
_{j}^{-}e^{-ik_{j}z}d\left( k_{j}z\right)  \label{eqFj}
\end{equation}
for the forward waves, and 
\begin{equation}
-c_{j}\frac{\partial B_{j}}{\partial z}+\frac{\partial B_{j}}{\partial t}+%
\frac{c_{j}}{2ik_{j}}\nabla _{\bot }^{2}B_{j}-\kappa _{j}B_{j}=-i\frac{b_{j}%
}{k_{j}}\frac{1}{2\pi }\int_{-\pi }^{\pi }\left[ p^{2}\right]
_{j}^{-}e^{ik_{j}z}d\left( k_{j}z\right)  \label{eqBj}
\end{equation}
for the backward waves.

Substituting the nonlinear terms (\ref{nonlinear}), only one of the
contributions survive, leading to a couple of equations for each frequency, 
\begin{eqnarray}
c_{0}\frac{\partial F_{0}}{\partial z}+\frac{\partial F_{0}}{\partial t}+%
\frac{c_{0}}{2ik_{0}}\nabla _{\bot }^{2}F_{0}-\kappa _{0}F_{0} &=&-i\frac{%
b_{0}}{4}F_{1}^{2},  \nonumber \\
-c_{0}\frac{\partial B_{0}}{\partial z}+\frac{\partial B_{0}}{\partial t}+%
\frac{c_{0}}{2ik_{0}}\nabla _{\bot }^{2}B_{0}-\kappa _{0}B_{0} &=&-i\frac{%
b_{0}}{4}B_{1}^{2}  \label{eq0}
\end{eqnarray}
for the fundamental wave, and 
\begin{eqnarray}
c_{1}\frac{\partial F_{1}}{\partial z}+\frac{\partial F_{1}}{\partial t}+%
\frac{c_{1}}{2ik_{1}}\nabla _{\bot }^{2}F_{1}-\kappa _{1}F_{1} &=&-i\frac{%
b_{1}}{2}F_{1}^{\ast }F_{0},  \nonumber \\
-c_{1}\frac{\partial B_{1}}{\partial z}+\frac{\partial B_{1}}{\partial t}+%
\frac{c_{1}}{2ik_{1}}\nabla _{\bot }^{2}B_{1}-\kappa _{1}B_{1} &=&-i\frac{%
b_{1}}{2}B_{1}^{\ast }B_{0}  \label{eq1}
\end{eqnarray}
for the subharmonic.

Besides the existence of the counter-propagating waves, the cavity also
imposes a condition which relates the amplitudes of forward and backward in
given points of the resonator. The fields obey the following boundary
conditions in the cavity: in $z=0$%
\begin{eqnarray}
F_{0}\left( x,y,0,t\right) &=&\sqrt{{\cal R}_{0}}B_{0}\left( x,y,0,t\right) +%
{\cal T}_{0}Y,  \nonumber \\
F_{1}\left( x,y,0,t\right) &=&\sqrt{{\cal R}_{1}}B_{1}\left( x,y,0,t\right) ,
\label{boundary1}
\end{eqnarray}
and in $z=L$%
\begin{eqnarray}
B_{0}\left( x,y,L,t\right) &=&\sqrt{{\cal R}_{0}}e^{-i\theta
_{0}}F_{0}\left( x,y,L,t\right) ,  \nonumber \\
B_{1}\left( x,y,L,t\right) &=&\sqrt{{\cal R}_{1}}e^{-i\theta
_{1}}F_{1}\left( x,y,L,t\right)  \label{boundary2}
\end{eqnarray}
where $L$ is the length of the cavity, ${\cal T}_{0}$ is the transmitivity
of the boundary to the input wave (pump), with amplitude given by $Y$, $%
{\cal R}_{j}$ is the reflectivity of the field $p_{j}$ at the boundary.
Finally, $\theta _{j}$ are the detunings (frequency mismatch with respect to
the cavity), given by 
\begin{eqnarray}
\theta _{0} &=&2\pi m-2k_{0}L=\frac{\omega _{0}^{c}-\omega _{0}}{c_{0}/2L},
\label{det0} \\
\theta _{1} &=&2\pi n-2k_{1}L=\frac{\omega _{1}^{c}-\omega _{1}}{c_{1}/2L},
\label{det1}
\end{eqnarray}
being $\omega _{j}^{c}$ the frequency of the cavity (eigenmode) nearest to
the field frequency $\omega _{j}.$

The field evolution is described by the system of equations (\ref{eq0}) and (%
\ref{eq1}) together with (\ref{boundary1}) and (\ref{boundary2}). However,
the description can be greatly simplified after the introduction of the
following changes \cite{Lugiato88a}: 
\begin{eqnarray}
\tilde{F}_{0} &=&e^{\frac{1}{2L}\left( z-L\right) (\ln {\cal R}_{0}-i\theta
_{0})}F_{0}+\frac{\sqrt{{\cal T}_{0}}}{\sqrt{{\cal R}_{0}}}e^{i\frac{\theta
_{0}}{2}}\frac{1}{2L}\left( z-L\right) Y  \nonumber \\
\tilde{B}_{0} &=&e^{-\frac{z}{2L}(\ln {\cal R}_{0}-i\theta _{0})}B_{0}e^{i%
\frac{\theta _{0}}{2}}-\frac{\sqrt{{\cal T}_{0}}}{\sqrt{{\cal R}_{0}}}e^{i%
\frac{\theta _{0}}{2}}\frac{1}{2L}\left( z-L\right) Y  \nonumber \\
\tilde{F}_{1} &=&e^{\frac{1}{2L}\left( z-L\right) (\ln {\cal R}_{1}-i\theta
_{1})}F_{1},  \label{changes} \\
\tilde{B}_{1} &=&e^{-\frac{z}{2L}(\ln {\cal R}_{1}-i\theta _{1})}B_{1}e^{i%
\frac{\theta _{1}}{2}},  \nonumber
\end{eqnarray}

The great advantaje of the latter changes is that the boundary conditions
for these new fields are 
\begin{eqnarray}
\tilde{F}_{0}\left( 0,t\right) &=&\tilde{B}_{0}\left( 0,t\right) ,  \nonumber
\\
\tilde{F}_{1}\left( L,t\right) &=&\tilde{B}_{1}\left( L,t\right)
\label{boundnew}
\end{eqnarray}
which correspond to those of an ideal cavity with perfectly reflecting
boundaries. This fact will be used later for determining the unknow
longitudinal distribution.

Substituting the new amplitudes in the evolutions equations we find the new
system

\begin{align}
\frac{\partial \tilde{F}_{0}}{\partial t}+c_{0}\frac{\partial \tilde{F}_{0}}{%
\partial z}& =-c_{0}\frac{\left| \ln {\cal R}_{0}\right| +i\theta _{0}}{2L}%
\tilde{F}_{0}+c_{0}\frac{\left| \ln {\cal R}_{0}\right| +i\theta _{0}}{4L^{2}%
}\left( z-L\right) \frac{\sqrt{{\cal T}_{0}}}{\sqrt{{\cal R}_{0}}}Ye^{i\frac{%
\theta _{0}}{2}}+  \nonumber \\
& \frac{c_{0}}{2L}\frac{\sqrt{{\cal T}_{0}}}{\sqrt{{\cal R}_{0}}}Ye^{i\frac{%
\theta _{0}}{2}}-\frac{c_{0}}{2ik_{0}}\nabla ^{2}\tilde{F}_{0}-\kappa _{0}%
\tilde{F}_{0}-i\frac{b_{0}}{4}{\cal D}\left( z\right) \tilde{F}_{1}^{2}
\label{eqF02}
\end{align}
where we have defined the exponential term 
\begin{equation}
{\cal D}\left( z\right) =e^{\frac{1}{2L}\left( z-L\right) (\ln {\cal R}%
_{0}-i\theta _{0})}e^{\frac{-2}{2L}\left( z-L\right) (\ln {\cal R}%
_{1}-i\theta _{1})}.  \label{exp}
\end{equation}

Consider now two additional conditions: first, that the reflectivity at the
boundaries is high (weal losses), which is expressed as ${\cal R}%
_{0}\rightarrow 1$, and consequently ${\cal T}_{0}\rightarrow 0.$\ Second,
that the field frequencies are close to a resonant frequency of the cavity,
so that $\theta _{0}\rightarrow 0.$ Under these conditions, also known as
the mean-field limit in nonlinear optics, the exponential factor ${\cal D}%
\left( z\right) $ approaches unity, and the second term at the {\it l.h.s.}
of (\ref{eqF02}) can be neglected. Furthermore, in this limit the field
amplitudes $\tilde{F}_{j}$ and $\hat{B}_{j}$ approach to their real values, $%
F_{j}$ and $B_{j}$. The equation (\ref{eqF02}) takes then a simplified form,
which can be more conveniently written as 
\begin{equation}
\frac{\partial F_{0}}{\partial t}+c_{0}\frac{\partial F_{0}}{\partial z}%
=E-\gamma _{0}\left( 1+i\Delta _{0}\right) F_{0}+ia_{0}\nabla ^{2}F_{0}-i%
\frac{b_{0}}{4}F_{1}^{2}  \label{eqF03}
\end{equation}
where we have defined the new pump $E$, detuning $\Delta _{0}$, losses $%
\gamma _{0}$\ and diffraction $a_{0}$ parameters as, 
\begin{eqnarray}
E &=&\frac{c_{0}}{2L}\frac{\sqrt{{\cal T}_{0}}}{\sqrt{{\cal R}_{0}}}Ye^{i%
\frac{\theta _{0}}{2}}, \\
\Delta _{j} &=&\frac{\theta _{j}}{\left| \ln {\cal R}_{j}\right| +\frac{2L}{%
c_{j}}\kappa _{j}}, \\
\gamma _{j} &=&\frac{\left| \ln {\cal R}_{j}\right| c_{j}}{2L}+\kappa _{j},
\\
a_{j} &=&\frac{c_{j}}{2k_{j}}
\end{eqnarray}
where $j=0,1$. Defined in this way, the detuning is a quantity of the order
of 1. Also, $\gamma _{0}$ represents a measure of the linewidth of the
cavity modes.

We have performed the derivation for the forward component of the
fundamental wave. A similar procedure leads to the evolution equations for
the other components, 
\begin{eqnarray}
\frac{\partial B_{0}}{\partial t}+c_{0}\frac{\partial B_{0}}{\partial z}
&=&E-\gamma _{0}\left( 1+i\Delta _{0}\right) B_{0}+ia_{0}\nabla ^{2}B_{0}-i%
\frac{b_{0}}{4}B_{1}^{2},  \label{eqB02} \\
\frac{\partial F_{1}}{\partial t}+c_{1}\frac{\partial F_{1}}{\partial z}
&=&-\gamma _{1}\left( 1+i\Delta _{1}\right) F_{1}+ia_{1}\nabla ^{2}F_{1}-i%
\frac{b_{1}}{2}F_{1}^{\ast }F_{0},  \label{eqF12} \\
\frac{\partial B_{1}}{\partial t}+c_{1}\frac{\partial B_{1}}{\partial z}
&=&-\gamma _{1}\left( 1+i\Delta _{1}\right) B_{1}+ia_{1}\nabla ^{2}B_{1}-i%
\frac{b_{1}}{2}B_{1}^{\ast }B_{0}  \label{eqB12}
\end{eqnarray}

The equations still keep a explicit $z$ dependence. Consider now the new
boundary conditions, given by (\ref{boundnew}). These conditions differ from
that of the original fields in that they represent an ideal (lossless)
cavity, which allows to express the field inside the cavity by means of the
Fourier expansions 
\begin{eqnarray}
\hat{F}_{j}\left( {\bf r},t\right) &=&\sum_{n=-\infty }^{\infty
}p_{j}^{\left( n\right) }\left( x,y,t\right) \exp \left( i\frac{n\pi z}{L}%
\right) ,  \nonumber \\
\hat{B}_{j}\left( {\bf r},t\right) &=&\sum_{n=-\infty }^{\infty
}p_{j}^{\left( n\right) }\left( x,y,t\right) \exp \left( -i\frac{n\pi z}{L}%
\right)  \label{fourier}
\end{eqnarray}

From (\ref{fbmodes}), it follows that the total field with frequency $\omega
_{j}$ can be also written as 
\begin{equation}
P_{j}({\bf r},t)=\sum_{n=-\infty }^{\infty }2p_{j}^{\left( n\right) }\left(
x,y,t\right) \cos \left[ \left( k_{j}+\frac{n\pi }{L}\right) z\right]
e^{-i\omega _{j}t}+c.c.  \label{cos2}
\end{equation}
If we finally assume that the intermode spacing $c/2L$ is large, we can
consider that only the $n=0$ mode can be excited, and $F_{0}$ and $B_{0}$
became spatially uniform along the longitudinal axis of the cavity. In this
case, the field can be described by (\ref{cosmodes}), which is the usual
description for waves in acoustic resonators.

This last assumptions makes the dynamical evolution to be independent on $z$%
, and Eq. (\ref{eqF03}) take the final form 
\begin{eqnarray}
\frac{\partial p_{0}}{\partial t} &=&E-\gamma _{0}\left( 1+i\Delta
_{0}\right) p_{0}+ia_{0}\nabla ^{2}p_{0}-i\frac{b_{0}}{4}p_{1}^{2}, 
\nonumber \\
\frac{\partial p_{1}}{\partial t} &=&-\gamma _{1}\left( 1+i\Delta
_{1}\right) p_{1}+ia_{1}\nabla ^{2}p_{1}-i\frac{b_{1}}{2}p_{1}^{\ast }p_{0}
\label{eq1fin}
\end{eqnarray}
which are the equations (\ref{psg1}) given in the text.

\newpage

{\LARGE Figure captions: \vspace{2cm}}

Figure 1: Eigenvalue as a function of the perturbation wavenumber, for $%
\Delta _{1}=1$, $E=2.25$ (a), and for $\Delta _{1}=-1$, $E=1.35$ (b).{\LARGE 
\vspace{1cm}}

Figure 2: Schematic representation of the longitudinal cavity resonance
condition for negative detunings.{\LARGE \vspace{1cm}}

Figure 3: The linear stage of the evolution. It is shown the pressure
amplitude (left column) and the corresponding spectrum (right column) in the
transverse plane. Pictures were taken at times $t=1$ (a) and $t=5$ (b) The
parameters used in the integration are $\Delta _{1}=-2$, $\Delta _{0}=-1$, $%
E=2$, $\gamma _{0}=\gamma _{1}=1$, $a_{1}=0.001$ and $a_{0}=0.0005.${\LARGE 
\vspace{1cm}}

Figure 4: Nonlinear evolution and pattern selection. The same parameters as
in Fig.3 have been used. Pictures were taken at times $t=25$ (a), $t=750$
(b) and $t=1000$ (c).

\clearpage

\begin{figure}[b]
\begin{center}
\includegraphics[width=0.5\textwidth]{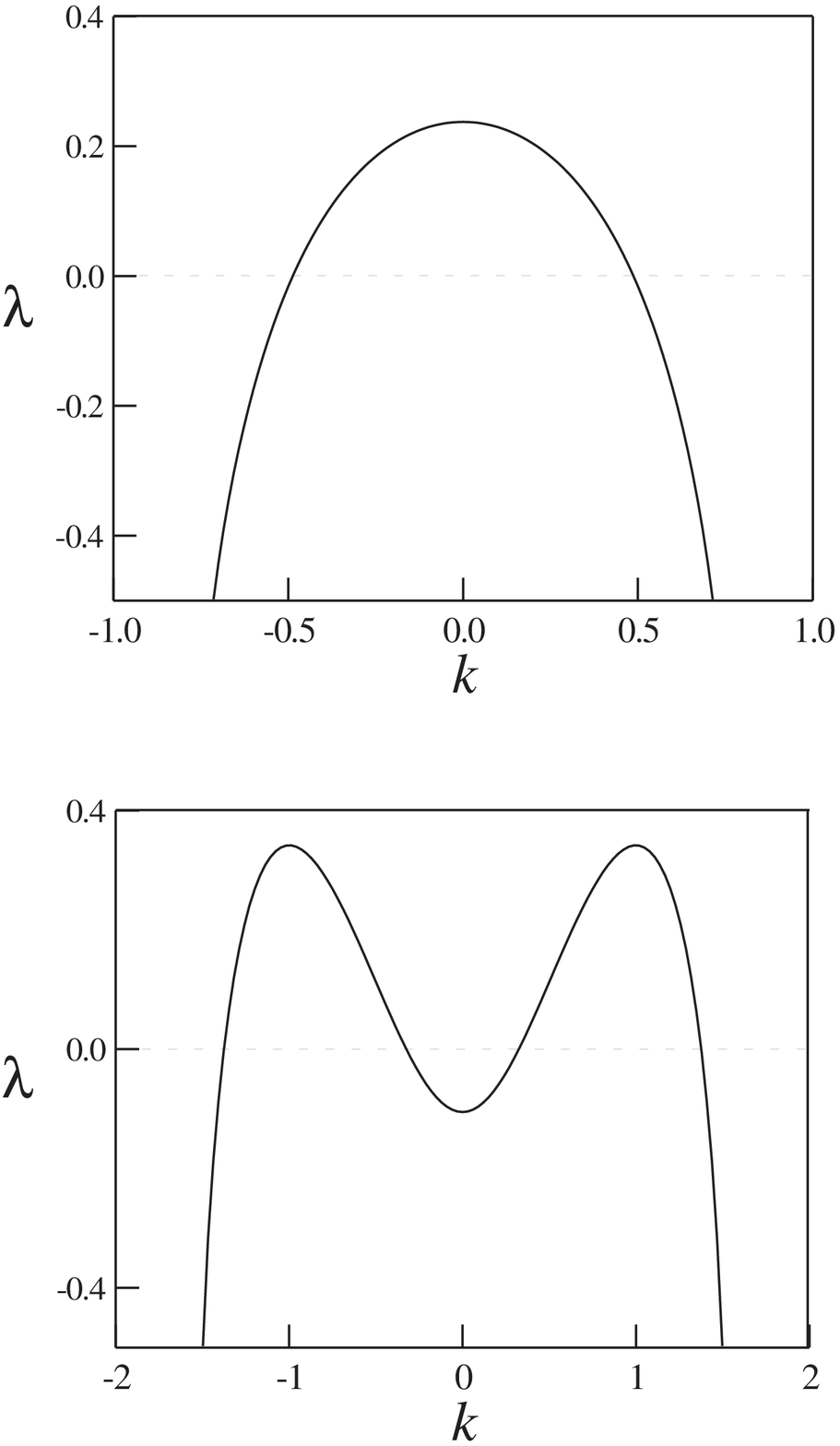}
\caption{{}}
\end{center}
\end{figure}
\clearpage

\begin{figure}[b]
\begin{center}
\includegraphics[width=0.5\textwidth]{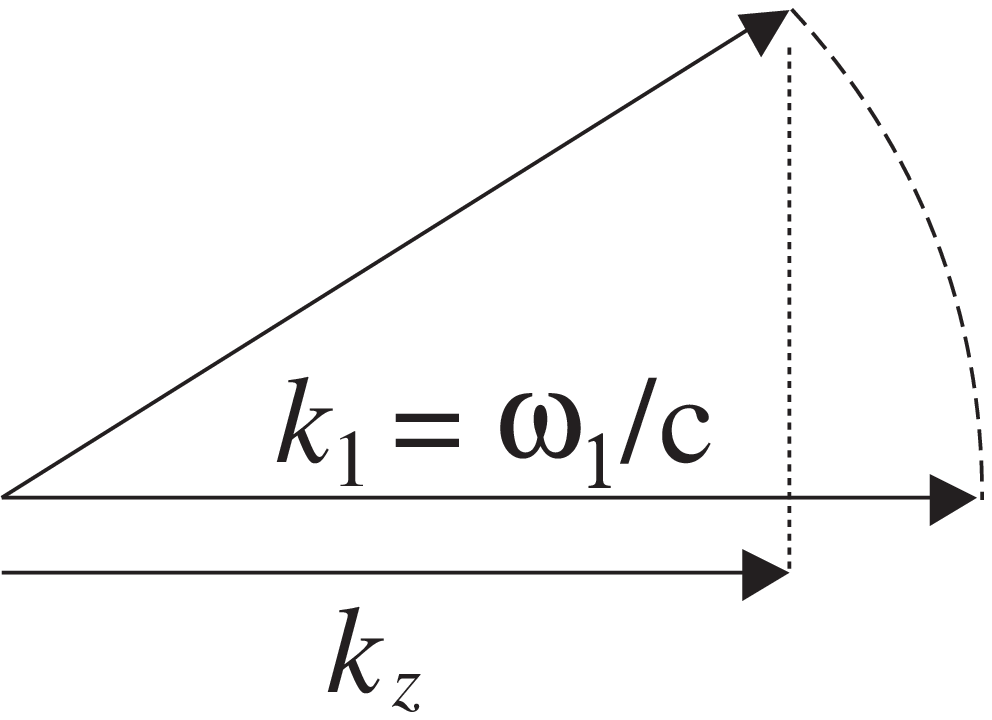}
\caption{{}}
\end{center}
\end{figure}

\clearpage

\begin{figure}[b]
\begin{center}
\includegraphics[width=0.5\textwidth]{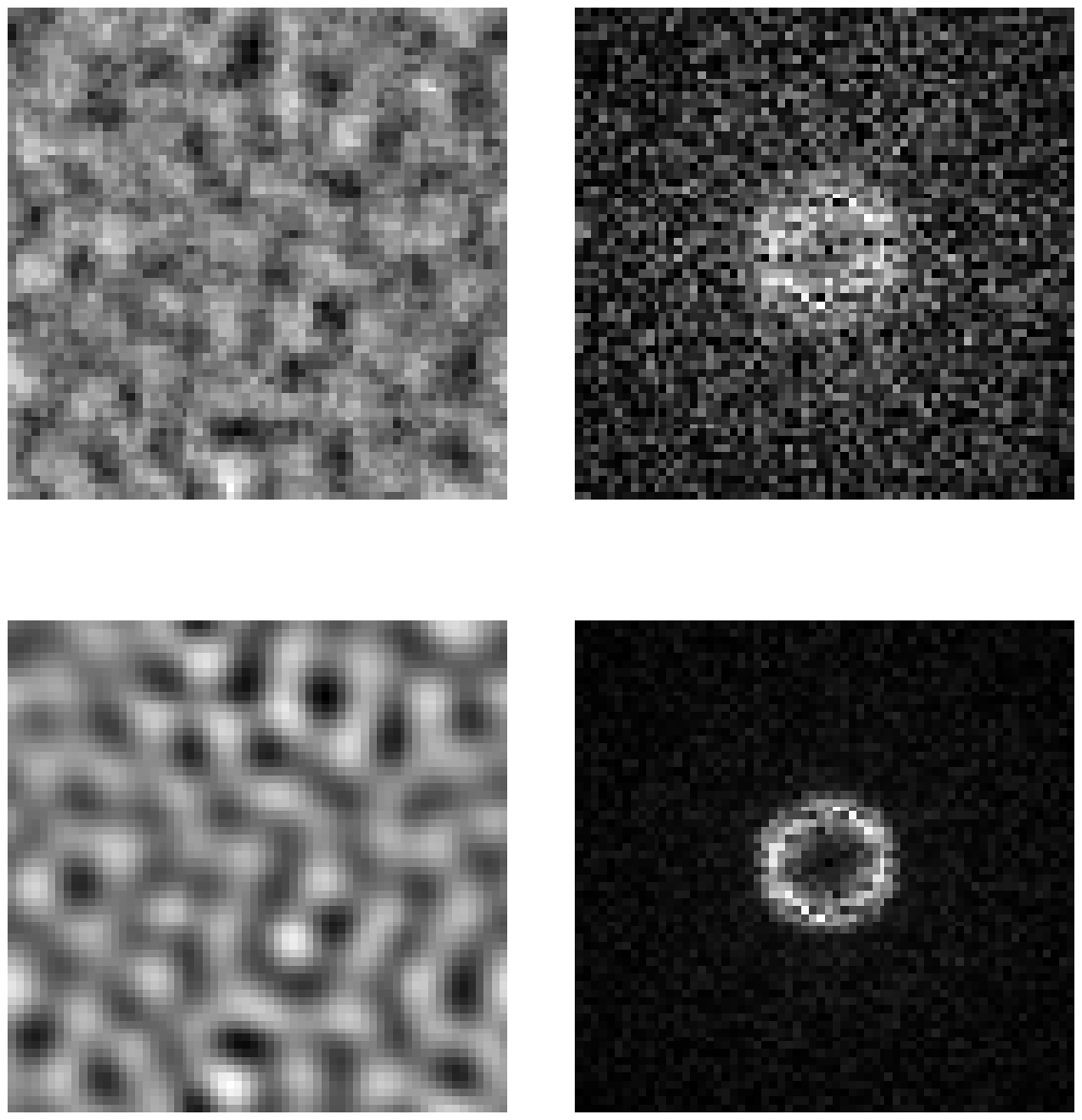}
\caption{{}}
\end{center}
\end{figure}

\clearpage

\begin{figure}[b]
\begin{center}
\includegraphics[width=0.5\textwidth]{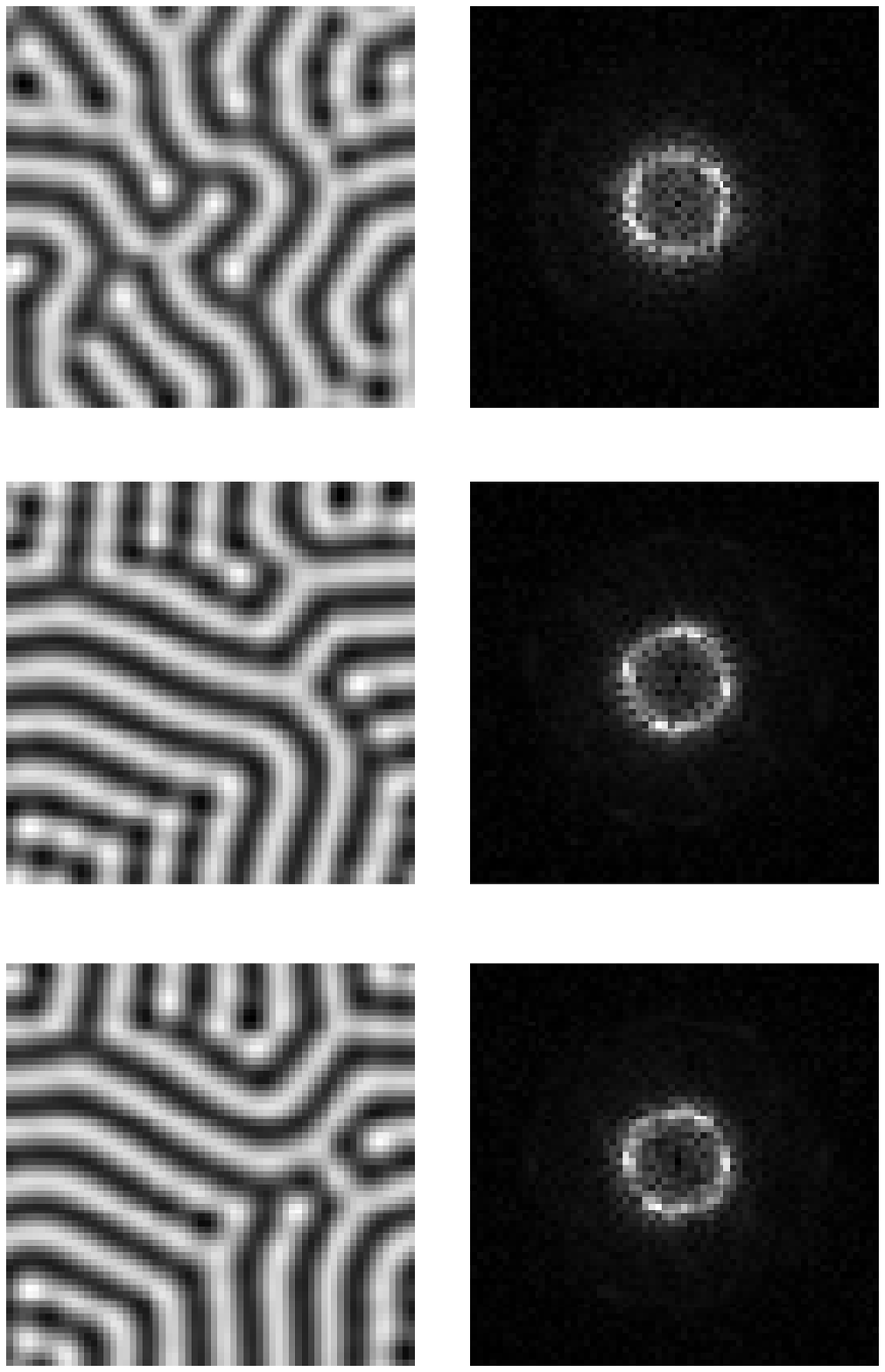}
\caption{{}}
\end{center}
\end{figure}

\end{document}